\documentclass[aps,prl,twocolumn,showpacs]{revtex4}
\usepackage{latexsym}
\usepackage{amsfonts}
\usepackage{amsmath}
\usepackage{bm}
\usepackage{amssymb}
\usepackage{graphicx}
\begin{document}
\title{Sticky obstacles to intramolecular energy flow}
\author{R. Pa\v{s}kauskas$^1$}
\email{rytis@gatech.edu}
\author{C. Chandre$^2$}
\author{T. Uzer$^1$}
\affiliation{$^1$ Center for Nonlinear Sciences, School of Physics,
Georgia Institute of Technology, Atlanta, Georgia 30332-0430, U.S.A.\\
$^2$ Centre de Physique Th\'eorique - CNRS, Luminy - Case 907, 13288 Marseille
cedex 09, France}

\date{\today}

\begin{abstract}
Vibrational energy flows unevenly in molecules, repeatedly going back and forth
between trapping and roaming. We  identify
bottlenecks between diffusive and chaotic behavior, and describe generic mechanisms of these transitions, 
taking the carbonyl sulphide molecule {OCS} as a case study. The
bottlenecks are found to be lower-dimensional tori; their bifurcations and unstable manifolds govern the
transition mechanisms.
\end{abstract}


\maketitle
Chemical reactions usually proceed through a complex choreography of
energy flow processes that deliver the needed vibrational energy to
the reactive mode. The manner and time in which energy travels
determine the outcome of the reaction and the properties of the
products. The conventional wisdom concerning this fundamental process
is that vibrational energy travels very fast and well before a reaction takes place, distributes itself
statistically among the modes of the molecule, assumed to resemble an
ensemble of coupled oscillators.
Reaction rate theories based on these assumptions~-­~known
collectively as statistical
theories~\cite{pechukas76}~-­~have been
vindicated in a number of chemical reactions. However, there is
increasing evidence that the approach to equilibrium usually proceeds more
slowly than predicted by statistical theories~\cite{chaosfocus05}~--~%
and it is also nonuniform, showing intriguing fits and
starts. This anomalous diffusion is caused by variety of phase space
structures, such as 
resonant islands or tori \cite{zasl05} that strongly
slow down the trajectories passing nearby~\cite{zasl05,keshavamurthy06} and 
therefore are said to be ``sticky''~\cite{perry94}. To date, the theories so
successfully applied in pioneering works~\cite{davis85,davis86,rice87,martens87,skodje88} to lower-dimensional systems have not been extended beyond two
degrees of freedom due to severe technical difficulties~
\cite{gillilan90,ezra,mikitotoda}.
\begin{figure}[!t]%
  \centerline{
    \includegraphics[width=8.5cm]{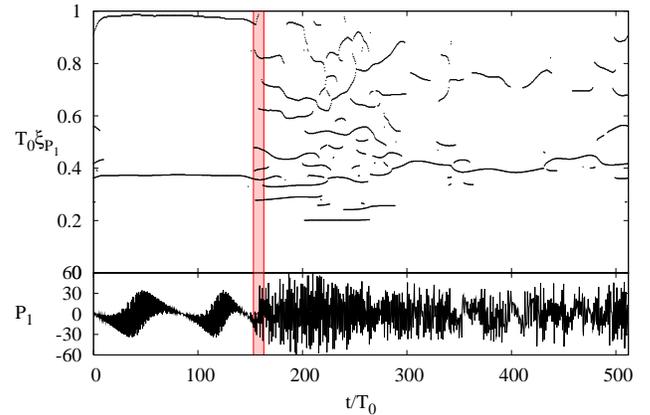}
  }
  \caption{\label{fig:Fig1}
    The generic behavior of chaotic trajectories in Hamiltonian systems involves
    substantial fraction of intermittent behavior. 
    The time-frequency analysis of a typical {OCS} trajectory (top panel) allows one to register the
    transition region (shaded band) and the frequencies $\xi$ of the
    regular motion, while the time series (lower panel) display the
    striking features of this abrupt change.
    $t$ is time (in units of $T_{0}=0.063\times10^{-12}$ s) and $\xi_{P_1}$ are the frequency ridges (in units of $T_0^{-1}$) in the time-frequency decomposition~\cite{xtel3} of $P_1(t)$.}%
\end{figure}
\begin{figure*}[!t]
\includegraphics*[scale=1.0]{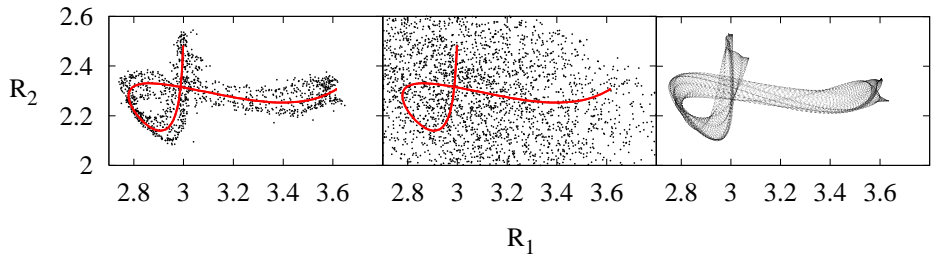}
\caption {Projections of the
  trajectory near a periodic orbit ${\mathcal O}_a$ (with period $T_0$), analyzed in
  Fig.~\ref{fig:Fig1}. The trajectory is represented in
  ($R_1$, $R_2$) plots, broken down into segments, corresponding to
  the {\em trapping stage} (left panel) and  chaotic stage (center
  panel). The bottleneck of transition from diffusion to hyperbolicity
  can be identified as a two-dimensional invariant torus (right
  panel.) The trajectory is sampled at fixed time intervals
  $T_0/2$. The orbit ${\mathcal O}_a$ is shown as a solid curve in the center.}
\label{fig:Fig2}
\end{figure*}

The {OCS}
molecule, an important player in greenhouse effect~\cite{turco80},
displays the slow and uneven relaxation to statistical equilibrium
mentioned above, despite its three strongly coupled degrees of
freedom. Models of {OCS} have served as a testbed for studying intramolecular dynamics
in the chaotic regime~\cite{carter82} and these classical findings have been confirmed in parallel quantal wave packet calculations ~\cite{gibson86}.
In this Letter, we investigate vibrational energy flow in the
{OCS} molecule using a Hamiltonian of the form ~\cite{foord75}
\begin{eqnarray}
  \label{eq:hamilt}
  H=T(R_1,R_2,\alpha,P_1,P_2,P_\alpha)+V(R_1,R_2,\alpha),
\end{eqnarray}
where $T$ is the standard kinetic energy of a rotationless triatomic molecule represented by two interatomic distances $R_1$ and $R_2$, and a bending angle $\alpha$ (with their canonically conjugate momenta $P_1$, $P_2$ and $P_\alpha$). The potential $V$ is fitted as
\begin{equation}
  V(R_1,R_2,\alpha)=\sum\limits_{i=1}^3 V_i(R_i)+V_I(R_1,R_2,R_3),
\end{equation}
where $R_3$ is the distance between O and S. The potential consists of Morse potentials $V_i$ for each diatomic pair
and
an interaction potential $V_I$ of the Sorbie-Murrell form~\cite{carter82}.
A rich mixture of chaotic and
regular dynamics is observed at energies close to dissociation~\cite{shch04}. The computations below were performed at 90\% of the
dissociation energy of the weaker bond. Trajectories in the vicinity of a specific periodic orbit with
elliptic normal stability are studied, focusing on their
escape to the chaotic
region, and identifying a generic mechanism of crossover from diffusion~\cite{laskar93}
to hyperbolicity and chaos. 

\begin{figure}
  \includegraphics{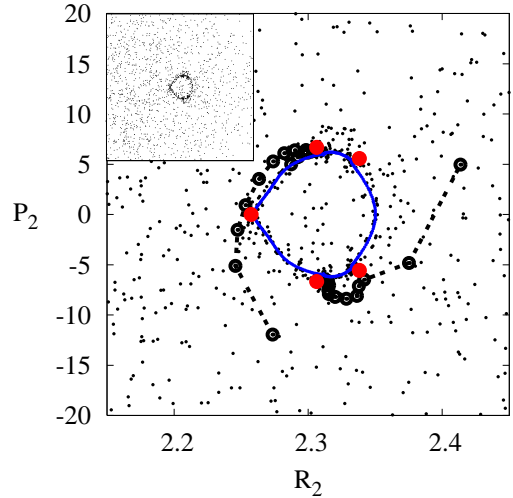}%[width=8.5cm]
  \caption{\label{fig:Fig3}
    Poincar\'e section of the trajectory near a periodic orbit ${\mathcal O}_a$,
    analyzed in Figs.~\ref{fig:Fig1}
    and~\ref{fig:Fig2}. The bottleneck (a two-dimensional torus) is a loop
    (blue) at the bifurcation point  (``D'' in
    Fig.~\ref{fig:Fig4}). The trajectory is trapped in the vicinity of a loop
    (which is clearly seen from the inset). The escape stage is shown
    as two ``tentacles,'' which extend along the unstable manifolds of a
    resonant periodic orbit (the five red dots around the center).
  }
\end{figure}

Figure~\ref{fig:Fig1} displays the time series
of such a trajectory, initially close to the periodic orbit ${\mathcal O}_a$ with period $T_0$ (see Ref.~\cite{shch04}).
Figure~\ref{fig:Fig2} shows 
salient features of capture (left panel) followed by an abrupt
transition to chaos (center panel).
An alternative view of the transition mechanism appears in Fig.~\ref{fig:Fig3} in terms of the  
Poincar\'e section $\Sigma:\,P_{\alpha}=0,\,\dot{P}_{\alpha}>0,\,\alpha\leq\pi$.
 A boundary, marking the crossover from diffusion to hyperbolicity, can be
 identified in terms of an invariant two-dimensional torus (right panel of
 Fig.~\ref{fig:Fig2}).
Normal bifurcations of two-dimensional tori turn out to be the key ingredients in the transition
mechanism to hyperbolicity, as will be shown below.
Generically, there are two
stages in the dynamics of the trajectory.
During the {\em trapping stage} (duration $t_{\mathrm{trap}}$), the trajectory
is close to (quasi-)periodic, following the unstable manifold of normally hyperbolic
tori with very small {\em positive} Lyapunov exponent
(in our case, $\lambda\simeq 10^{-2}$, thus explaining the observed trapping time
$t_{\mathrm{trap}}\sim \lambda^{-1}$).
During the {\em escape stage}  (duration $t_{\mathrm{esc}}$; the shaded band in
Fig.~\ref{fig:Fig1} and ``tentacles'' in Fig.~\ref{fig:Fig3}),
the trajectory follows the unstable manifolds of the periodic
orbit which is in 3:5 resonance with ${\mathcal O}_a$ (thick dots in Fig.~\ref{fig:Fig3}), with a 
significantly larger Lyapunov exponent, leading to a fast transition
to the chaotic region of phase space (center panel of Fig.~\ref{fig:Fig2}).
These two time scales usually satisfy $t_{\mathrm{trap}} \gg t_{\mathrm{esc}}$.
Observations of repeated trapping-escape-chaotic processes in
relatively short trajectory segments ($\sim10^{3}T_0$) provide
evidence that these effects are prevalent.
Dynamical systems
theory identifies structures with minimal hyperbolicity as key players in
describing long-term features of the chaotic component of an
attractor, and integral surfaces with small positive Lyapunov exponent
are candidates for the ``backbone'' of the dynamics. Normally hyperbolic invariant manifolds~\cite{HPS77}
 have recently been implicated in the symbolic dynamics and phase space partition of higher-dimensional chaotic Hamiltonian systems~\cite{llave:priv07,rytis07hl},
systems with small-dimensional saddles such as the
``Crossed Fields''~\cite{uzer02} 
and the Restricted Three Body
Problems~\cite{gomez04}. 

Using a
combination of trajectory diagnostic tools like Lyapunov 
maps~\cite{shch04,froe97b}, time-frequency analysis
\cite{xtel3}, and methods from the theory of
dynamical systems like periodic and
quasiperiodic orbit computations~\cite{simo98:effective,jorba01}, 
we relate the phenomenon of trapping to invariant structures in
phase space and to lower-dimensional invariant tori (with a relation to
their normal stability properties) in particular.
It is commonly assumed that in ``typical'' Hamiltonian systems with a large
number of degrees of freedom $N$, the relative measure of $N$-dimensional
invariant tori ($N$ local integrals) is either zero
or one~\cite{froeschle71}. The implication is 
that chaotic systems with large $N$ approach conditions of the
stochastic ansatz, and hence, the trapping phenomenon described above is insignificant. On the other
hand, it has been established recently that high order resonances form robust islands of
secondary structures with positive measure~\cite{haro00}. 

\begin{figure}[!t]
  \includegraphics{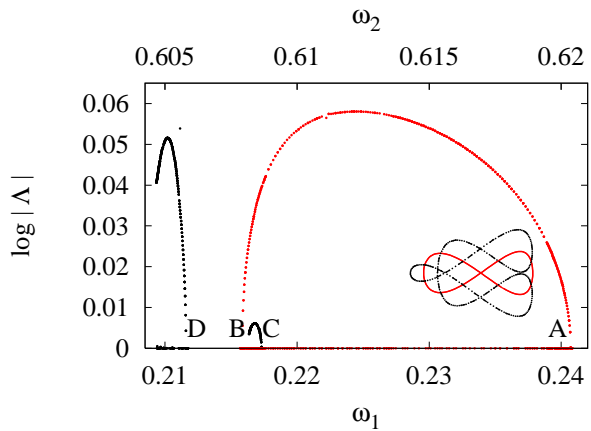}
  \caption{\label{fig:Fig4}
    Fine structure of invariant tori, scanned along the transition
    channel. The plot shows how Lyapunov exponents depend on the
    rotation number $\omega$. The points of frequency halving
    bifurcations (``A''-- ``D'') can be interpreted as
    bottlenecks of transition from diffusion to hyperbolicity.
    Red dots: family of loops arising from
    the periodic orbit ${\mathcal O}_a$. Black dots: 
    frequency halved loop, emerging at the bifurcation point ``A''.  
    Insets display ($R_1$,$P_1$) projections of loops near the bifurcation point ``A''. Red: loop with elliptic normal stability and $\omega=\omega_1\approx0.24067$.
    Black: loop with hyperbolic normal stability and $\omega=\omega_2=(\omega_1+1)/2\approx0.62033$.
  }
\end{figure}

In order to identify bottlenecks of transition from diffusion to chaos, we monitor the
progress of invariant phase space structures along the transition
channel using rotation numbers. The results are summarized in Fig.~\ref{fig:Fig4}, which is central to
understanding this transition. In a trapping region around the
elliptic periodic orbit ${\mathcal O}_a$ (left panel of Fig.~\ref{fig:Fig2}),
the rotation numbers are obtained from the frequency map
analysis~\cite{laskar93} on the surface of
section. It can be characterized by a single
$\omega_{\mathrm{trap}}\approx 0.60556$, implying that a
two-dimensional torus is the relevant invariant structure
in the trapping process. Having computed a family of two-dimensional tori, parametrized by rotation
numbers $\omega$, it is evident that
$\omega_{\mathrm{trap}}$ places the torus on the hyperbolic branch of the bifurcation diagram
represented in Fig.~\ref{fig:Fig4}. This implies that the
escape is mediated by manifolds of a torus with hyperbolic normal stability. The duration of the trapping stage is approximately 150
returns on $\Sigma$, and is consistent with the maximal Lyapunov exponent $\lambda<0.05$.  
Processes associated with the escape from the
trapping region can be better understood by analyzing
the tangent space of the elliptic periodic orbit ${\mathcal O}_a$ that locally has
the structure of a direct product (center $+$ center)
${\mathbb T}\times{\mathrm I}_1\times{\mathbb T}\times{\mathrm I}_2$, with the periodic
orbit at the origin. The elements of the two intervals
 ${\mathrm I}_i\subset{\mathbb R}$  are rotation numbers $\omega_i$, which are not unique in general: The choice is fixed by requiring $\lim_{\mu \rightarrow 0}\omega_i=\omega^{0}_i$, where $\mu$ is a measure of the torus and
$\omega^{0}_i$ are stability angles of the elliptic periodic orbit
${\mathcal O}_a$ ($\omega_1^0=0.24500633$ and
$\omega_2^0=0.37046872$). The Poincar\'{e} map induces rotations on ${\mathbb T}$, 
$r_{\omega_1}\times{\mathrm 1}\times
r_{\omega_2}\times{\mathrm 1}$, where $r_{\omega}$ is a rotation on ${\mathbb T}$ with the rotation
number $\omega$. Partial (or complete) resonances are
determined by one (or two) resonance conditions
$n\omega_1+m\omega_2+k=0$, where $(n,m,k)$ are integers such that $|n|+|m|+|k|>0$. The most striking
trapping effects are observed for partial resonances of the type
${\mathbb T}\times{\mathrm I}_1\times\{0\}\times\{0\}$, and
$\{0\}\times\{0\}\times{\mathbb T}\times{\mathrm I}_2$. Choosing either of
the two situations, a resonance channel has been constructed by finding the two-dimensional
invariant tori for $\omega_i\in{\mathrm I}_i$. In order to find these
tori we consider the  Poincar\'{e} map ${\mathcal F}_{\Sigma}: \Sigma\mapsto\Sigma$.
Tori may have hyperbolic normal linear stability,
therefore a search for them cannot rely on methods exploiting
``stickiness'' properties. The sections of two-dimensional invariant tori
are one-dimensional closed curves (called hereafter
``loops''). 
We consider loops as discretizations of  
$\gamma:\,{\mathbb T}\mapsto\Sigma$ (with periodic boundary condition
 $\gamma(s)=\gamma(s+1)$) and require that the Poincar\'{e} map
 ${\mathcal F}_{\Sigma}$, restricted to the loop is equivalent to a rigid rotation 
$r_{\omega}$. This translates into an invariance condition:
\begin{equation}
  \label{eq:loop}
  {\mathcal F}_{\Sigma}( \gamma (s) ) = \gamma( s + \omega ).
\end{equation}
Equation~(\ref{eq:loop}) is solved using damped Newton iterations for the
Fourier coefficients of $\gamma(s)$. The linear stability properties of the
loop are determined by $(\Lambda,\psi)$, solutions of the
generalized eigenvalue problem: 
\begin{equation}\label{eq:Dloop}
  D{\mathcal F}_{\Sigma}(s)\psi(s) = \Lambda\psi(s+\omega).
\end{equation}
Equation~(\ref{eq:Dloop}) has a one-dimensional kernel, which we eliminate
using singular value decomposition. 
The initial data for the Newton iterations $\gamma_0(s)$ and
$\omega$ were obtained using one of the following two methods~:
The first method uses the trapping region of the trajectory (see
Fig.~\ref{fig:Fig1}). We estimate $\omega$ using Fourier-like
methods~\cite{laskar93}, and truncate the continued
fraction expansion of $\omega=[a_1,a_2,\ldots]$ before the first large
$a_i$ so that $\omega_{0}=P/Q$. 
Then we take
sequences of trapping region data every $Q$ iterations and combine them to
obtain $\gamma_0(s)$. A refined value of $\omega$ can be
estimated by minimizing
$| {\mathcal F}_{\Sigma}\circ\gamma_0-\gamma_0\circ r_{\omega}|$. 
The second method combines continuation in $\omega$ with the direct product
structure in the neighborhood of the periodic orbit.  The surface of
section derivative $D{\mathcal F}_{\Sigma}$ at the periodic orbit has two pairs of
complex eigenvalues $\exp{[\pm\iota\omega_i^0]}$, $i=1,2$.  The
eigenvectors define mutually skew orthogonal symplectic vector spaces
$V_i={\mathbb R}^2$. It is assumed here that the linear approximation is effective in the neighborhood of the periodic
orbit.

The set of two-dimensional tori is found to be discontinuous at the
gaps in Fig.~\ref{fig:Fig4} due to complete
resonances (periodic orbits) and secondary invariant structures.
Normal stability is typically
elliptic for small $|\omega-\omega_i^0|$. We identify
the two-dimensional invariant torus at the period doubling bifurcation point as a bottleneck of a
given resonance channel. The rationale follows from the theory of
dynamical systems: Beyond the bifurcation point at $\omega=\omega^c$, the normal stability
changes to hyperbolic. This change affects trajectories passing by
its neighborhood. One recurrent observation is that the continued fraction expansion of
bifurcation rotation numbers has a tail composed of small integers
(see Tab.~\ref{tab:study2_bif}). This feature is reminiscent of the observation that the continued fraction expansion of the frequency of the last invariant torus in generic Hamiltonian systems with two degrees of freedom is noble (with a tail of ones) in many situations~\cite{davis85}. 
\begin{table}
  \begin{tabular}{l|l|l}\hline
    $\omega$           &   value        &   Cont. frac.\\\hline
    $\omega^{c}_{A}$ & 0.240711317575 & [4, 6, 2, 11, 5, 5\ldots] \\
    $\omega^{c}_{B}$ & 0.215852976389 & [4, 1, 1, 1, 2, 1, 1, 1, 1, 2, 1, 1\ldots]\\
    $\omega^{c}_{C}$ & 0.608654398762 & [1, 1, 1, 1, 4, 45, 1, 1, 1, 1\ldots]\\
    $\omega^{c}_{D}$ & 0.605804087926 & [1, 1, 1, 1, 6, 3, 2, 2, 1\ldots]\\
  \end{tabular}  
  \caption{\label{tab:study2_bif} Rotation numbers of the two-dimensional invariant tori at the bifurcation points A, B, C, D shown in Fig.~4.}
\end{table}


The reliability of the numerical solution can be tested by
examining its Floquet multipliers, given by Eq.~(\ref{eq:Dloop}). 
An exact
solution consists of a set of complex numbers with up to
three different absolute values: $1$, $\Lambda$,
$1/\Lambda$. Significant variation from these values signals an unreliable 
solution.

In conclusion, our findings indicate that trapping and escape are mediated by the
same sequence of events, and an approximate boundary, which separates
trapped and chaotic behavior, can be found in analogy with the
boundaries that separate reactants from products in Transition State
Theory~\cite{pechukas76}, where sharply defined phase space
structures~\cite{pollak78,MacKay1,MacKay2,uzer02} play this role.

In a broader context, our work forms yet another stimulus to
reconsider the relevance of local integrals and
partial resonances in realistic, chaotic Hamiltonian systems with many
degrees of freedom. Here, we have explained a
paradoxical situation, namely that integral surfaces with {\it  positive}
Lyapunov exponents (i.e., \ not ``sticky'') can trap chaotic trajectories.
Widespread observations of repeated trapping-escape-chaotic processes in
short trajectory segments provide evidence that these effects are
generic and occurring frequently in many settings ranging from plasmas to celestial mechanics.

This research was partially supported by the US National Science Foundation. C.C. acknowledges support from Euratom-CEA (contract EUR~344-88-1~FUA~F).
%\bibliography{../bibtex/rytis}
%\bibliographystyle{apsrev}
\newcommand{\noopsort}[1]{} \newcommand{\printfirst}[2]{#1}
  \newcommand{\singleletter}[1]{#1} \newcommand{\switchargs}[2]{#2#1}

\end{document}